\documentclass[letterpaper,10pt,conference]{ieeeconf} 

\usepackage{amsmath}
\usepackage{amssymb}
\usepackage{mathtools}  

\usepackage{graphicx}
\usepackage{xcolor}
\usepackage{epstopdf}

\usepackage{algorithmic}
\usepackage[ruled,vlined]{algorithm2e}

\usepackage{caption}
\usepackage{subcaption}
\usepackage{float}

\usepackage{tabularx,booktabs}

\usepackage{cite}
\usepackage{balance}
\usepackage{flushend}
\usepackage{tikz}

\newtheorem{theorem}{\textbf{Theorem}}

\newtheorem{assumption}{\textbf{Assumption}}

\DeclareGraphicsExtensions{.pdf,.png,.jpg,.jpeg,.eps}
\newcommand{\norm}[1]{\left\lVert#1\right\lVert}
\usepackage[export]{adjustbox}
\IEEEoverridecommandlockouts  

\usepackage{comment}
\usepackage{color}
\usepackage{epsfig} 
\graphicspath{{Figures/}}
\usepackage{textcomp}
\usepackage{flafter}
\usepackage{sidecap}
\usepackage{comment}
\usepackage{psfrag}
\usepackage{nomencl}
\usepackage{stackengine}
\usepackage{dsfont}
\makenomenclature

\usepackage{fancyhdr}

\newtheorem{problem}{\textbf{Problem}}
\newtheorem{prop}{\textbf{Proposition}}
\newtheorem{defn}{\textbf{Definition}}

\title{{{\bf Resilient Interval Observer-Based Control for Cooperative Adaptive Cruise Control under FDI Attack}}}

\author{Parisa Ansari Bonab, Elisabeth Andarge Gedefaw, Mohammad Khajenejad
\thanks{Parisa Ansari Bonab, Elisabeth Andarge Gedefaw,  and Mohammad Khajenejad are with the Department of Mechanical Engineering at the University of Tulsa, Tulsa, Oklahoma, 74104, (\texttt{emails: \{paa3545, eag8758, mok7673@utulsa.edu\}}).}
\thanks{This work is partially supported by National Science Foundation grant CPS:CRII-451042.}%
}
\date{September 2024}

\begin{document}

\maketitle
\thispagestyle{empty}
\pagestyle{empty}

\begin{abstract}
Connectivity in connected and autonomous vehicles (CAVs) introduces vulnerability to cyber threats such as false data injection (FDI) attacks, which can compromise system reliability and safety. To ensure resilience, this paper proposes a control framework combines a nonlinear controller with an interval observer for robust state estimation under measurement noise. The observer bounds leader's states, while a neural network-based estimator estimates the unknown FDI attacks in real time. These estimates are then used to mitigate FDI attack effects and maintain safe inter-vehicle spacing. The proposed approach leverages an idea of interval observer-based estimation and merges model-based and learning-based methods to achieve accurate estimations and real-time performance. MATLAB/Simulink results confirm resilient tracking, precise FDI attack estimation, and robustness to noise, demonstrating potential for real-world CACC applications under cyberattacks, disturbance, and bounded measurement noise.

\end{abstract}

\providecommand{\keywords}[1]
{
  \small	
  \textbf{\textit{Energy efficiency}} #1
}



\section{Introduction} \label{ssec:intro}
According to the National Highway Traffic Safety Administration, 42,514 fatalities occurred on U.S. roads in 2022 due to motor vehicle crashes \cite{national2024overview}. Studies show that human error is responsible for nearly 90\% of all accidents, while mechanical failures account for only about 2\% \cite{singh2015critical}. To address these challenges, automated vehicles (AVs) use onboard sensors to perceive the driving environment and enhance road safety. Vehicles equipped with wireless communication operate as connected vehicles (CVs), enabling vehicle-to-vehicle (V2V) communication to exchange information such as position, velocity, and acceleration, thereby improving situational awareness and traffic efficiency \cite{kazemi2018learning}.

Integrating automation and connectivity leads to connected and automated vehicles (CAVs), which combine local sensing with V2V communication for more reliable perception and decision-making \cite{electronics13101875, shladover2018connected}. One of the fundamental functionalities of AVs is adaptive cruise control (ACC), which adjusts a vehicle’s velocity to maintain a desired following distance. However, ACC suffers from limitations such as string instability at short headways and delayed response times comparable to human drivers, both of which can negatively impact traffic flow and safety \cite{ciuffo2021requiem, brunner2022comparing}.

Extending ACC through V2V communication forms cooperative adaptive cruise control (CACC), a key feature of CAVs. CACC allows vehicles to coordinate more precisely with their predecessors, improving stability, responsiveness, and spacing efficiency \cite{moradipari2024benefits, boddupalli2022resilient}. Despite these advantages, the reliance on wireless communication makes CACC vulnerable to cyber threats, particularly false data injection (FDI) attacks, where adversaries alter transmitted information to disrupt vehicle coordination \cite{niroumand2024security, siddiqi2024multichain, 10354110}.

\emph{\textbf{Literature Review.}} Over the past several years, numerous studies have investigated FDI attacks targeting CACC systems \cite{biroon2021false,wolf2020securing,van2017analyzing,zhao2021detection,amoozadeh2015security,bezemskij2017detecting,petrillo2017collaborative,sargolzaei2016machine}. In \cite{biroon2021false}, a partial differential equation (PDE)-based approach was developed to identify FDI attacks in vehicle platoons by comparing the attack signatures under normal and compromised scenarios. Similarly, \cite{wolf2020securing} examined multiple mitigation strategies and found that employing an attack-resilient controller offered the highest potential for improving system security, although no explicit detection or control algorithm for CACC was presented. The work in \cite{van2017analyzing} further emphasized that detection alone is insufficient, demonstrating that combining a resilient controller with an effective detection scheme achieves superior mitigation.

In \cite{zhao2021detection}, a cloud-based detection algorithm was proposed to accurately identify and isolate compromised vehicles, but it lacked a complementary control strategy to counteract detected attacks. Another line of research explored connectivity management as a defense strategy, disabling communication links during attacks to revert CACC operation to traditional ACC. In contrast, \cite{petrillo2017collaborative} introduced a consensus-based control mechanism that preserves CACC functionality by continuously monitoring vehicles for anomalies and isolating malicious agents, though it did not incorporate an explicit detection framework.

A learning-based approach using neural networks was proposed in \cite{sargolzaei2016machine} to detect and estimate FDI attacks in CACC. The combination of neural estimation and control effectively mitigated attack effects; however, the stability of the resulting closed-loop system was not analyzed. 

In \cite{ansari2024secure}, a secure Lyapunov-based controller was developed for CACC under FDI attacks and external disturbance. The study introduced a control and estimation framework that combines model-based and learning-based techniques to achieve both high accuracy and real-time performance. The proposed controller maintained reliable tracking of the lead vehicle even under adversarial conditions, with stability established using Lyapunov theory and validation performed through both simulation and experimental results. While this work achieved accurate real-time estimation and control, it relied on point estimation of the leader’s position and did not explicitly account for bounded measurement noise in the sensing process.

In practical CACC, measurement noise is inevitable due to imperfect radar, lidar, and wireless measurements. These bounded uncertainties can significantly degrade the estimation accuracy and, consequently, the safety and stability of the overall system could not properly addressed. Traditional point observers estimate a single state trajectory and often fail to guarantee robustness in the presence of bounded measurement noise. To overcome this limitation, interval observers have been introduced as a powerful alternative. An interval observer estimates the upper and lower bounds of system states rather than a single value, ensuring that the true states always remain within these intervals despite noise and model uncertainties \cite{10886809, 10384243, khajenejad2025guaranteedprivatecontrol, 11284789, KHAJENEJAD2025101328, 11107747}.

 \emph{\textbf{Contribution.}} This paper extends the secure control on CACC framework originally proposed in \cite{ansari2024secure}.  In this extended work, we address the limitation of the original work by introducing an interval observer-based framework. This observer explicitly considers bounded measurement noise in the radar-based measurements of the leader’s position and velocity. The use of an interval observer allows for the estimation of upper and lower bounds of the leader’s position, thereby providing a more realistic and noise-resilient estimation structure that better reflects real-world conditions where sensor uncertainty is inevitable. The proposed approach ensures that estimation errors remain bounded while maintaining resilience against FDI attacks and external noise and disturbance.



\section{Preliminaries}\label{preliminaries}
For a matrix $M \in \mathbb{R}^{n \times p}$, the $i$-th row and $j$-th column entry is denoted by $M_{ij}$. We define $M^{\oplus} \triangleq \max(M, 0_{n \times p})$, 
$M^{\ominus} \triangleq M^{\oplus} - M$, 
and $|M| \triangleq M^{\oplus} + M^{\ominus}$, 
which represents the element-wise absolute value of $M$. 
Furthermore, $M^{d}$ denotes the diagonal matrix containing only the diagonal entries of the square matrix 
$M \in \mathbb{R}^{n \times n}$, while 
$M^{nd} \triangleq M - M^{d}$ contains only the off-diagonal elements. 
The matrix $M^{m} \triangleq M^{d} + |M^{nd}|$ is known as the ``Metzlerized'' matrix\footnote{The Metzler matrix is defined as a matrix whose off-diagonal elements are nonnegative.}. 
Additionally, we use $M \succ 0$ and $M \prec 0$ (or $M \succeq 0$ and $M \preceq 0$) to indicate that $M$ is positive definite and negative definite (or positive semi-definite and negative semi-definite), respectively. 
All vector and matrix inequalities are element-wise inequalities, and the zero and ones matrices of size $n \times p$ are denoted by $0_{n \times p}$ and $\mathbf{1}_{n \times p}$, respectively.
\begin{defn}[Interval]\label{defn
} $\mathcal{I} \triangleq[\underline{z}, \overline{z}] \subset \mathbb{R}^n$ is an n-dimensional interval defined as the set of all vectors $z \in \mathbb{R}^{n_z}$ that satisfy $\underline{z} \leq z \leq \overline{z}$. Interval matrices can be defined in a similar way. \end{defn}
\begin{prop}
\cite[Definition 4]{Khajenejad2025}[Correct Interval {Framers} and Framer Errors]\label{defn:framers}
For any generic $n$-dimensional continuous-time dynamic  system $\mathcal{G}$, with state $x(t)$ and bounded process and measurement noise signals $w(t)\in \mathcal{W}$ and $v(t) \in \mathcal{V}$, the signals/sequences $\overline{x},\underline{x}: {\mathbb{R}_{\geq 0}} \to \mathbb{R}^n$ are called upper and lower framers for the system, if 
$\underline{x}(t) \leq x(t) \leq \overline{x}(t), \ \forall t \in {\mathbb{R}_{\geq 0}}, {\forall w(t) \in \mathcal{W},\forall v(t) \in \mathcal{V}}$. Moreover, the dynamical system $\hat{\mathcal{G}}$ is called a framer system for $\mathcal{G}$, if its states/trajectory construct framer signals for the sates/trajectory of $\mathcal{G}$.
\end{prop}
\begin{prop}\cite[Definition 5]{Khajenejad2025}[$L_1$ Optimal Interval Observer]\label{defn:L_1}
An interval framer  $\hat{\mathcal{G}}$ is said to be $L_1$ optimal, if it minimizes the $L_1$-gain of the framer error ($\varepsilon(t)\triangleq \overline{x}(t)-\underline{x}(t)$) system $\tilde{\mathcal{G}}$, defined below.
\begin{align}\label{eq:L1_Def}
\|\tilde{\mathcal{G}}\|_{\ell_1} \triangleq \sup_{\|\delta\|_{\ell_1}=1} \|\varepsilon\|_{\ell_1}  
\end{align}
where $\|v\|_{\ell_1} \triangleq \int_0^\infty \|v(t)\|_1 dt$ denote the $\ell_{1}$ signal norms for $\nu \in \{\varepsilon, \delta\}$. The terms $\varepsilon_t$ and $\delta(t)=\delta\triangleq \begin{bmatrix} \delta_w^\top & \delta_v^\top \end{bmatrix}^\top$ represent the framer error and the combined noise signals, respectively, with $\delta_w \triangleq\overline{w}-\underline{w}$ and $\delta_v\triangleq\overline{v}-\underline{v}$.
\end{prop}
\section{Problem Statement}\label{Problem_Statement}
\subsection{Mathematical Model of CACC under FDI attacks and Measurement Noise} \label{cacc_model} 
We consider a string of CACC-equipped vehicles communicating with one another, where each vehicle receives data from its immediate leader. The transmitted information primarily includes the leader’s control signal. In addition to communication, each following vehicle employs an onboard radar sensor to measure the velocity and position of its leader; these measurements are subject to noise, which is considered in the model. It is assumed that the vehicle string is homogeneous and that all vehicles in the platoon are identical in dynamics. The dynamic model represents the real dynamics of an experimental vehicle, obtained through system identification and validation in a controlled experimental setup, as explained in detail in \cite{ansari2024secure}. The dynamics of the following vehicle are described as follows
\begin{equation}\label{system_dynamics_1}
\begin{cases}
\dot{x}_{i}(t) = v_{i}(t),\,\\
\dot{v}_{i}(t) = -a_iv_{i}(t) + b_iu_i(t)+d_i(t),
\end{cases}
\end{equation}
where $i\in \{2,\cdots n\}$ represents the index of the follower vehicle, $n$ denotes the total number of vehicles in the platoon, and $i-1$ corresponds to the immediate leader of vehicle. The presented equations describe the dynamics of the $i^{th}$ follower relative to its leader $i-1$. In Equation \eqref{system_dynamics_1}, $a_i\in \mathbb{R}$ and $b_i\in \mathbb{R}$ are the constant parameters which were obtained from experimental analysis in \cite{ansari2024secure}. Also $x_i\in\mathbb{R}$, $v_i\in\mathbb{R}$, $u_i\in\mathbb{R}$, and $d_i\in\mathbb{R}$ represent the position, velocity, control input, and external disturbance, respectively. 
\begin{assumption} \label{disturbance_follower}
The disturbance signal $d_i(t)$ is continuous and bounded by known lower and upper bunds satisfying:$$\underline{d}_i(t) \leq d_i(t) \leq  \overline{d}_i(t), \quad \forall t.$$
\end{assumption}

\noindent \textbf{FDI Attacks Representation.}
FDI attacks are injected into the communication network of connected vehicles, causing the follower vehicle to receive a corrupted control signal from its leader. Such malicious interference can compromise system stability, leading to collisions within the vehicle platoon. To quantify the impact of these attacks, the attack function is defined in the following 
\begin{equation}\label{fdieq}
\pi_i(u_{i-1}(t)) \triangleq u_{i-1}(t) + f_{i-1}(t),
\end{equation}
where $\pi_i\in\mathbb{R}$ is the attack function, $u_{i-1}$ is the leader control command and $f_{i-1}(t)\in\mathbb{R}$ is the bounded, unknown, continuous, and time-varying FDI attack.

\begin{assumption}\label{Attack_Function}
$f_{i-1}(t)$ is assumed to be bounded and differentiable such that $\lvert f_{i-1}(t) \rvert \leq \overline{f}_{i-1}$, where $t \geq t_0$ and $\overline{f}_{i-1}$ is a positive constant.
\end{assumption}

\noindent\textbf{Measurement Noise.}
As discussed earlier, the follower vehicle employs a radar sensor to measure the velocity and position of its leader. These measurements are affected by bounded noise signal whose distribution is unknown. Since the leader’s position plays a crucial role in the controller design, particularly in computing the inter-vehicle distance error, an interval observer is utilized to accurately estimate it in the presence of the bounded measurement noise.\\ 

\vspace{-.2cm}
\noindent \textbf{Follower's Perspective.} Accordingly, the leader’s dynamics, as perceived from the follower’s perspective, are expressed:
\begin{align}\label{system_dynamics_2}
\begin{array}{rl}
\dot{x}_{i-1}(t) &= v_{i-1}(t),\\
\dot{v}_{i-1}(t) &= -a_{i-1}v_{i-1}(t) + b_{i-1}\Bar{u}_{i-1}(t)+d_{i-1}(t),\\
y_{i-1}(t)&=v_{i-1}(t)+\theta_{i-1}(t),
\end{array}
\end{align}
where $a_{i-1},b_{i-1}\in \mathbb{R}$ are the constant parameters same as $a_i$ and $b_i$, as the dynamic models for all the vehicles are similar. In addition, $x_{i-1}\in\mathbb{R}$, $v_{i-1}\in\mathbb{R}$, $\Bar{u}_{i-1} \triangleq u_{i-1} +f_{i-1}$ , $d_{i-1}\in\mathbb{R}$, and $\theta_{i-1}(t)\in\mathbb{R}$ represent the position, velocity, corrupted control input, external disturbance, and measurement noise of the leader from the view of follower, respectively. Furthermore, For determining the accuracy of the control signal estimation, an estimation error for the control signal, $\tilde{u}_{i-1}:[t_0,\infty)\to\mathbb{R}^{n_i}$, is defined as
\begin{equation}
\tilde{u}_{i-1}(t) \triangleq  u_{i-1}(t) - \hat{u}_{i-1}(t),   
\end{equation}
where $\hat{u}_{i-1}$ and ${u}_{i-1}$ are the estimated and actual control signal of the leader, respectively. Next, defining $\Bar{u}_{i-1}(t) \triangleq u_{i-1}(t) +f_{i-1}(t)$ and $\hat{u}_{i-1}(t) \triangleq \Bar{u}_{i-1}(t) - \hat{f}_{i-1}(t)$ yields
\begin{equation}\label{u-bar}
\tilde{u}_{i-1}(t) = u_{i-1}(t) - \overline{u}_{i-1}(t) +\hat{f}_{i-1}(t), 
\end{equation}
where $\hat{f}_{i-1}$ is an estimation of the FDI attack from the follower's perspective.

To measure the accuracy of the FDI attack estimation, the estimation error for the FDI attack, $\tilde{f}_{i-1}:[t_0,\infty)\to\mathbb{R}^{n_i}$ is: 
\begin{equation}\label{estimate-error}
\tilde{f}_{i-1}(t) \triangleq f_{i-1}(t) - \hat{f}_{i-1}(t).
\end{equation}
Finally, the estimation error for the leader's position is: 
\begin{equation}\label{estimate-error-leader-position}
\tilde{x}_{i-1}(t) \triangleq \overline{x}_{i-1}(t) - \underline{x}_{i-1}(t),
\end{equation}
where $\overline{x}_{i-1}(t)$ and $\underline{x}_{i-1}(t)$ are to-be-obtained upper and lower bounds of the leader's position. 

Moreover, we assume the following: 
\begin{assumption} \label{disturbance_noise_leader}
The disturbance and noise signals are continuous and bounded by known lower and upper bund as $\underline{d}_{i-1}$, $\overline{d}_{i-1}$, $\underline{\theta}_{i-1}$, and $\overline{\theta}_{i-1}$, further, $x_{i-1}(0) \in [\overline{x}_{i-1}(0), \underline{x}_{i-1}(0)]$ with known $\overline{x}_{i-1}(0)$ and $\underline{x}_{i-1}(0)$.
\end{assumption}
\begin{assumption}\label{ass:bounded_distance}\cite{patre2008asymptotic}
The desired distance between vehicles, $x_{d_i}$, as well as its first and second derivatives are bounded ($x_{d_i},\dot{x}_{d_i},\Ddot{x}_{d_i}\in\mathcal{L}_{\infty}$) by positive known constants.
\end{assumption}
\noindent \textbf{Objectives.} The primary objective is to develop a resilient control framework that ensures safe inter-vehicle spacing under the influence of FDI attacks, bounded measurement noise, and external disturbances. In CACC, the follower vehicle relies on real-time control signals from the leader; however, these signals can be compromised through malicious manipulation, potentially resulting in unsafe spacing or collisions. To address this vulnerability, a secondary objective is to design a real-time FDI attack estimation mechanism capable of identifying and mitigating the effects of injected data on the control system. In addition, since bounded measurement noise is considered in the sensor readings used to obtain the leader’s velocity and position, the final objective is to design an interval observer capable of estimating the upper and lower bounds of the leader’s position and use the midpoint of them as an approximation for the leader's position from the follower's perspective. 


The considered problem can be stated as follows.
\begin{problem}\label{prob:resilient_estimation}
 Design a resilient control and estimation framework that ensures the desired inter-vehicle distance is maintained under FDI attacks, disturbances, and bounded measurement noise. Specifically, the controller should drive the distance error, $e_{i} \triangleq x_{i}-x_{i-1}+D_{i-1}+x_{d_i}$, to zero, the FDI attack estimator should minimize the attack estimation error $\tilde{f}_{i-1}(t)$, and the interval observer should be designed to accurately estimate the leader’s position despite measurement noise and minimizes the estimation error $\tilde{x}_{i-1}(t)$. 
\end{problem}
\section{Proposed Solution} \label{solution}
To address Problem \ref{prob:resilient_estimation}, a nonlinear Lyapunov-based controller, an FDI attack estimator, and an interval observer are proposed as follows.
\subsection{Proposed Control Strategy \& FDI Attack Estimation}\label{controller-section}
Inspired by our previous work~\cite{ansari2024secure}, we propose the following control strategy:
\begin{equation}\begin{aligned}\label{control-signal}
u_i(t) &\triangleq \frac{a_i}{b_i}v_{i}(t)-\frac{a_{i-1}}{b_{i-1}}v_{i-1}(t)+\overline{u}_{i-1}-\hat{f}_{i-1}(t)-\frac{1}{b_i}\Ddot{x}_{d_i}(t) \\& -\frac{\alpha_i}{b_i}r_i(t) +\frac{\alpha^2_i}{b_i}e_i(t)-\frac{1}{b_i}e_i(t)-\frac{K_{1_i}}{b_i}r_i(t),
\end{aligned}\end{equation}
where $K_{1_i}\in\mathbb{R}_{>0}$ is a gain specified by the user. Furthermore, to estimate the attack signal, we leverage a
neural network-based estimation algorithm~\cite{chakraborty2017control}, which returns the following estimation of the FDI attack signal:
\begin{equation}\label{estimate-NN}
\hat{f}_{i-1} \triangleq \hat{W}_i^T\sigma(\hat{V}_i^T\delta_i),
\end{equation}
where $\hat{W}_i\in\mathbb{R}^{(n_i+1)\times n_i}, \hat{V}_i\in\mathbb{R}^{(n_i+1)\times n_n}$ represent the estimated ideals weights: 
\begin{align}\label{update-law1}
\begin{array}{rl}
\dot{\hat{W}}_i &= proj(\Gamma_{1_i}\sigma(\hat{V}_i^T\delta_i)\phi_i),\\
\dot{\hat{V}}_i &=  proj(\Gamma_{2_i}\phi_i\delta_i\hat{W}_i^T\sigma'(\hat{V}_i^T\delta_i)),
\end{array}
\end{align}
while $\delta_i \overset{\Delta}{=} [1,\phi_i^T]^T$, $\phi_i \overset{\Delta}{=} b_ir_i$, and ``proj" denotes a Lipschitz continuous projection operator defined in \cite{9720959}, and $\Gamma_{1_i}, \Gamma_{2_i} \in\mathbb{R}^{n_i\times n_i}$ are definite positive matrices (More details are provided in Appendix and \cite{9720959}).
\subsection{Observer Design} \label{observer}
We first rewrite the dynamics in \eqref{system_dynamics_2} as the following equivalent representation:
\begin{align}\label{Leader_dynamic_1}
\hspace{-.3cm}  \begin{array}{rl}
\dot{X}_{i-1}(t) &\hspace{-.3cm}= \hspace{-.1cm}A_{i-1}X_{i-1}(t)\hspace{-.1cm} +\hspace{-.1cm} B_{i-1}\bar{u}_{i-1}(t)\hspace{-.1cm} +\hspace{-.1cm} W_{i-1}d_{i-1}(t), \\
y_{i-1}(t) &\hspace{-.3cm}= C_{i-1}X_{i-1}(t) + V_{i-1}\theta_{i-1}(t),
\end{array}
 \end{align}
where $
X_{i-1}(t) \triangleq 
\begin{bmatrix}
x_{i-1}(t) \\
v_{i-1}(t)
\end{bmatrix}$,
$A_{i-1} \triangleq 
\begin{bmatrix}
0 & 1 \\
0 & -a_{i-1}
\end{bmatrix} 
$, $B_{i-1} \triangleq 
\begin{bmatrix}
0 &
b_{i-1}
\end{bmatrix}^\top,
W_{i-1} \triangleq 
\begin{bmatrix}
0 &
1
\end{bmatrix}^\top$, $C_{i-1} \triangleq 
\begin{bmatrix}
0 & 1
\end{bmatrix}$, and $V_{i-1} = 1$. Then,  
to estimate the upper and lower bounds of the leader’s position and stimulated by our previous work~\cite{Khajenejad2025}, we propose the following interval observer:
\begin{equation}\label{Observer_Design}
\begin{aligned}
\hspace{-.3cm}\left\{
\begin{array}{l}
\underline{\dot{Z}}_{i-1}(t) = M_x^{\uparrow} \underline{Z}_{i-1}(t) - M_x^{\downarrow} \overline{Z}_{i-1}(t)
+ M_w^{\oplus} \underline{d}_{i-1} \\\hspace{1.25cm}- M_w^{\ominus} \overline{d}_{i-1} 
 + M_v^{\ominus} \overline{\theta}_{i-1} - M_v^{\oplus} \underline{\theta}_{i-1} 
\quad \\\hspace{1.25cm} + M_u \bar{u}_{i-1}(t) + (M_x N + L) y_{i-1}(t), \\
\dot{\overline{Z}}_{i-1}(t) = M_x^{\uparrow} \bar{Z}_{i-1}(t) - M_x^{\downarrow}\underline{Z}_{i-1}(t)
 + M_w^{\oplus} \overline{d}_{i-1} \\\hspace{1.25cm} - M_w^{\ominus} \underline{d}_{i-1}
 + M_v^{\ominus} \underline{\theta}_{i-1}  - M_v^{\oplus} \overline{\theta}_{i-1} 
\quad \\\hspace{1.25cm} + M_u \overline{u}_{i-1}(t) + (M_x N + L) y_{i-1}(t), \\
\underline{X}_{i-1}(t) = \underline{Z}_{i-1}(t) + (NV)^{\ominus} \overline{\theta}_{i-1} - (NV)^{\oplus} \underline{\theta}_{i-1} \\\hspace{1.25cm} + N y_{i-1}(t), \\
\overline{X}_{i-1}(t) = \overline{Z}_{i-1}(t) + (NV)^{\ominus} \underline{\theta}_{i-1} - (NV)^{\oplus} \overline{\theta}_{i-1} \\\hspace{1.25cm} + N y_{i-1}(t),
\end{array}
\right.
\end{aligned}
\end{equation}
where $M_x\triangleq TA-LC$, $M_x^{\uparrow}\triangleq M_x^{d}+ {M_x^{nd}}^{\oplus}$, $M_x^{\downarrow}\triangleq {M_x^{nd}}^{\ominus}$, $M_w\triangleq TW$, $M_v\triangleq M_xN+L$, and $M_u\triangleq TB$. In addition, $\overline{Z}_{i-1}(t),\underline{Z}_{i-1}(t)$ are ``auxiliary" framers, and the observer gains $L,T,N$ are designed through the following proposition to guarantee that \eqref{Observer_Design} constructs an ISS and $\mathcal{L}_{1}$-optimal interval observer for \eqref{Leader_dynamic_1}.  
\begin{prop}\label{prop:obs_stab}
($L_1$-Optimal \& ISS 
{Observer Design}, \cite{Khajenejad2025}, Theorem 3). Suppose $X=Q^{-1}\tilde{X}, \forall X \in \{L,{T},N\}$, where $(Q,L,{T},N)$ is an optimal solution to 
\eqref{eq:LP}. Then, \eqref{Observer_Design} is an $L_1$-Optimal and ISS interval observer for \eqref{Leader_dynamic_1} and consequently for \eqref{system_dynamics_2}. 
\begin{align}\label{eq:LP}
\hspace{-0.2cm}\begin{array}{l}
\min\limits_{\substack{\{\gamma, \Delta,Q,\Omega,\Gamma,\tilde{L},\tilde{N},\tilde{T},\tilde{M}_x,N,\Phi,\tilde{W}^p,\tilde{W}^n,
\tilde{L}^p,\tilde{L}^n,N_V^p,N_V^n, \tilde{N}^p,\tilde{N}^n,\\ \tilde{T}^p,\tilde{T}^n,\tilde{M}_x^{d},\tilde{M}_x^{nd},\tilde{M}_x^{d,p},\tilde{M}_x^{d,n},\tilde{M}_x^{nd,p},\tilde{M}_x^{nd,n}\}}} \gamma \\
\text{s.t.} \  \mathbf{1}^\top_{n} [\Delta \ \Gamma \ \Phi ] < \gamma   \mathbf{1}^\top_{{n}_w+m+n_v}, \\
 \quad \ \ {\mathbf{1}^\top_{n}}\Omega < -\mathbf{1}^\top_{n}, \ \text{and} \ \{\mathbf{C}\} \ \text{holds, where}
\end{array}\hspace{-0.4cm}
\end{align}
\begin{align}\label{eq:conditions}
&\nonumber \quad \mathbf{C}=\\
&\begin{cases}
 \Delta=\tilde W^p\!+\!\tilde W^n\!\\
 \tilde{T}W=\tilde W^p-\tilde W^n,\ NV = N_{v}^p-N_{v}^n,\\ 
 \Phi = \tilde{L}^p_v+\tilde{L}^n_v,\\
 \tilde{L}V=\tilde{L}^p_v-\tilde{L}^n_v,\  \tilde{N}V=\tilde{N}^p_v-\tilde{N}^n_v, \\
 \Omega =\tilde{M}^\textnormal{d}_z+\tilde{M}^{\textnormal{nd},p}_z+\tilde{M}^{\textnormal{nd},n}_z\\
 \tilde S=\tilde S^p-\tilde S^n, \forall S \in \{T,L,N\},\\
 \tilde{M}_x=\tilde{M}^\textnormal{d}_x+\tilde{M}^{\textnormal{nd}}_x,\\
\tilde{M}^{\textnormal{nd}}_x=\tilde{M}^{\textnormal{nd},p}_x-\tilde{M}^{\textnormal{nd},n}_x, \ \tilde{M}^{\textnormal{d}}_x=\tilde{M}^{\textnormal{d},p}_x-\tilde{M}^{\textnormal{d},n}_x,\\
 \tilde{M}_x=\tilde{A} -\tilde{L}C, \ \tilde{M}^\textnormal{d}_x=\mathrm{diag}(\tilde{M}_x),\\
 \tilde{A} =  (\tilde{T}^p A_p - \tilde{T}^n A_n)^{nd},\\
 \tilde{T}=Q\!-\!\tilde{N} C,\
 \gamma \!>\! 0, Q \in \mathbb{D}^n_{>0}, \tilde Y^\nu \geq 0, \forall \nu \in \{p,n\},\\
 \forall Y\in \{W,N,L,T,L_v,N_v,M_x,M^\textnormal{d}_x,M^{\textnormal{nd}}_x\}.
\end{cases}
\end{align}
\end{prop}
As a result of Proposition \ref{prop:obs_stab}, the interval observer \eqref{eq:conditions}, returns stable intervals containing the leader's position, $[\underline{x}_{i-1}(t),\overline{x}_{i-1}(t)] \ni x_{i-1}(t)$, whose midpoint can be used as the estimation of the leader's position from the follower's perspective:
\begin{equation} \label{eq:leader_position}
    \hat{x}_{i-1}(t)= \frac{\bar{x}_{i-1}(t)+\underline{x}_{i-1}(t)}{2}.
\end{equation}
\section{Stability Analysis} 
 In this section, we show that the proposed control policy in combination with the NN-based FDI attack estimator and the interval observers stabilizes the closed-loop dynamics, i.e., derives the distance error to zero despite FDI attack, external disturbance, and bounded measurement noise.
\begin{theorem}
Suppose Assumptions \ref{disturbance_follower}--\ref{ass:bounded_distance} hold. Then, the nonlinear control policy \eqref{control-signal} along with the FDI attack estimator in \eqref{estimate-NN} ensures uniformly ultimately bounded stability of the dynamics in vehicle \eqref{system_dynamics_1}, when the leader's position is estimated using \eqref{eq:leader_position} as well as the interval observer \eqref{Observer_Design}. 
\end{theorem}\label{stability}
\begin{proof}
For simplicity, the time dependence $(t)$ is dropped. Consider a radially unbounded, positive definite, and continuously differentiable candidate Lyapunov function 
$V_{L_i}:\mathbb{R}^3 \times [0,\infty) \to \mathbb{R}_{\geq 0}$:
\begin{equation}\label{lyapunov-analysis}
V_{L_i} = \frac{1}{2}e^2_i+\frac{1}{2}r^2_i+H_i,
\end{equation}
where $e_{i}:[t_0, \infty) \to \mathbb{R}$ denotes the ``distance error":
\begin{equation}\label{error}
e_{i} \triangleq x_{i}-x_{i-1}+D_{i-1}+x_{d_i},
\end{equation}
while $D_{i-1}\in\mathbb{R}$ is the length of lead vehicle, and $x_{d_i}\in\mathbb{R}$ is the desired distance between vehicles. Moreover, $r_{i}$ denotes an ``auxiliary error" satisfying
\begin{equation} \label{r_error}
r_{i} \triangleq \dot{e}_{i} + \alpha_i{e_{i}},
\end{equation}
where $\alpha_i\in\mathbb{R}_{>0}$, is a user-specified known gain. Furthermore, $H_i:[t_0, \infty) \to \mathbb{R}_{\geq0}$ is defined as 
\begin{equation}
H_i \triangleq \frac{1}{2}tr(\tilde{W}^T_i\Gamma_{1_i}^{-1}\tilde{W}_i) + \frac{1}{2}tr(\tilde{V}_i^T\Gamma_{2_i}^{-1}\tilde{V}_i),
\end{equation}
where $\tilde{V}_i=V_i-\hat{V}_i$ and $\tilde{W}_i=W_i-\hat{W}_i$ are the inner and outer NN weight errors, respectively. Since $\tilde{W}_i$ and $\tilde{V}_i$ are bounded, $H_i$  is bounded by $\lvert H_i \rvert \leq H_{i,max}$ with some known $H_{i,max}\in\mathbb{R}_{>0}$. Furthermore, let $p_i \triangleq [e^T_i, r^T_i]^T,\psi_{1_i} \triangleq \frac{1}{2}\norm{p_i}^2$, and $\psi_{2_i} \triangleq \norm{p_i}^2$. 
It is immediate to see from \eqref{lyapunov-analysis} that the following inequality holds:
\begin{equation}\label{eq:Lyap_bound}
\psi_{1_i} \leq V_{L_i} \leq \psi_{2_i} + H_{i,max},
\end{equation}
while taking the derivative of \eqref{lyapunov-analysis} yields
\begin{align}\label{lyapunov-derivative}
\begin{array}{rl}
\dot{V}_{L_i} &= e_i\dot{e}_i+r_i\dot{r}_i \\& -tr(\tilde{W}_i\Gamma_{1i}^{-1}\dot{\hat{W}}_i)-tr(\tilde{V}_i\Gamma_{2i}^{-1}\dot{\hat{V}}_i).
\end{array}
\end{align}
On the other hand, taking the derivative of both sides of \eqref{r_error} and substituting \eqref{error} returns:
\begin{equation}\label{r-dot1}
\dot{r}_i = \Ddot{x}_i-\Ddot{x}_{i-1}+\Ddot{x}_{d_i}+\alpha_i\dot{e}_i.
\end{equation}
Replacing $\Ddot{x}_i$ and $\Ddot{x}_{i-1}$ and \eqref{u-bar} into \eqref{r-dot1} produces 
\begin{equation*}\begin{aligned}
\dot{r}_i = & -a_iv_i+b_iu_i+d_i+a_{i-1}v_{i-1}\\& -b_i\Bar{u}_{i-1}+b_i f_{i-1}-d_{i-1} +\Ddot{x}_{d_i}+\alpha_i\dot{e}_i,
\end{aligned}\end{equation*}
which combined with \eqref{r_error} and \eqref{control-signal} results in
\begin{equation}\label{final-r-dot_1}
\dot{r}_i = b_i\tilde{f}_{i-1}-K_{1_i}r_i-e_i+d_i-d_{i-1}.
\end{equation}
Substituting \eqref{r_error} and \eqref{final-r-dot_1} into \eqref{lyapunov-derivative} yields
\begin{align*}
\dot{V}_{L_i} = &-\alpha_ie_i^2+r_i(b_i\tilde{f}_{i-1}-K_{1_i}r_i+d_i-d_{i-1}) \\& -tr(\tilde{W}_i\Gamma_{1i}^{-1}\dot{\hat{W}}_i)-tr(\tilde{V}_i\Gamma_{2i}^{-1}\dot{\hat{V}}_i),
\end{align*}
which combined with \eqref{taylor-series} (cf. Appendix) results in
\begin{align}\label{phi-sub}
\nonumber\dot{V}_{L_i} = &-\alpha_ie_i^2+(\tilde{W}_i^T\sigma(\hat{V}_i^T\delta_i)+\hat{W}_i^T\sigma'(\hat{V}_i^T\delta_i)\tilde{V}_i^T\delta_i)\phi_i \\ &+r_iN_{n_i}+r_i(-K_{1_i}r_i+d_i-d_{i-1}) \\
\nonumber &-tr(\tilde{W}_i\Gamma_{1i}^{-1}\dot{\hat{W}}_i)-tr(\tilde{V}_i\Gamma_{2i}^{-1}\dot{\hat{V}}_i),
\end{align}
where $\phi_i\triangleq b_ir_i$.
Designing the weights of NN, $\dot{\hat{W}}_i$, $\dot{\hat{V}}_i$ through \eqref{update-law1} to remove the term $(\tilde{W}_i^T\sigma(\hat{V}_i^T\delta_i)+\hat{W}_i^T\sigma'(\hat{V}_i^T\delta_i)\tilde{V}_i^T\delta_i)\phi_i$ from the right hand side of \eqref{phi-sub}: 
\begin{equation}\begin{aligned}\label{before-young}
\dot{V}_{L_i} = &-\alpha_ie_i^2+ r_iN_{n_i} +r_i(-K_{1_i}r_i+d_i-d_{i-1}),
\end{aligned}\end{equation}
where $N_{n_i}$ is given in \eqref{eq:Nni}. By applying Young's Inequality and any choice of $\varepsilon_j >0, j \in \{1,2,3\}$, the positive terms in the right hand side of \eqref{before-young} can be upper bounded as:
\begin{align*}
\begin{array}{c}
r_iN_{n_i} \leq \frac{1}{2\varepsilon_0}r^2_i+\frac{\varepsilon_0}{2}N^2_{n_i}, \
r_id_i \leq \frac{1}{2\varepsilon_1}r^2_i+\frac{\varepsilon_1}{2}d^2_i, \\
r_id_{i-1} \leq \frac{1}{2\varepsilon_2}r^2_i+\frac{\varepsilon_2}{2}d^2_{i-1}, 
\end{array}
\end{align*}
which after plugging in \eqref{before-young} yields: 
\begin{align}\label{after-young}
\nonumber\dot{V}_{L_i} \leq & -\alpha_ie^2_i + \frac{1}{2\varepsilon_0}r^2_i+ \frac{\varepsilon_0}{2}N^2_{n_i} -K_{1_i}r^2_i+\frac{1}{2\varepsilon_1}r^2_i +\frac{\varepsilon_1}{2}d^2_i \\& 
+\frac{1}{2\varepsilon_2}r^2_i + \frac{\varepsilon_2}{2}d^2_{i-1}. 
\end{align}
On the other hand, Assumption \ref{disturbance_follower} implies $|d_i| \leq \tilde{d}_i \leq \tilde{d}_i \triangleq \max(|\overline{d}_i|,|\underline{d}_i|)$, and similarly for $|d_{i-1}|$. Using this and \eqref{after-young}:
\begin{align}\label{like-terms}
\dot{V}_{L_i} \leq   -\alpha_ie^2_i 
- (K_{1_i} - \frac{1}{2\varepsilon_0} - \frac{1}{2\varepsilon_1}- \frac{1}{2\varepsilon_2} )r^2_i + \varphi_i,
\end{align}
where $\varphi_i \triangleq  \frac{\varepsilon_0}{2}\Bar{N}_{n_i}^2 +\frac{\varepsilon_1}{2}\tilde{{d}_i}^2 +\frac{\varepsilon_2}{2}\tilde{d}_{i-1}^2$.

Finally, by choosing proper $(K_{1,i},\varepsilon_0,\varepsilon_1,\varepsilon_2)$ such that $K_{1_i} >  \frac{1}{2\varepsilon_0} +\frac{1}{2\varepsilon_1}+\frac{1}{2\varepsilon_2}$ holds, as well as defining $\alpha_{1_i} \triangleq K_{1_i} - \frac{1}{2\varepsilon_0} - \frac{1}{2\varepsilon_1}- \frac{1}{2\varepsilon_2}>0$ and $\alpha_{2_i} \triangleq \min\{\alpha_i,\alpha_{1_i}\}>0$, it follows from \eqref{lyapunov-analysis}, \eqref{eq:Lyap_bound}, and \eqref{like-terms} that: 
\begin{equation}\label{bounded-lyapunov}
\dot{V}_{L_i} \leq -\frac{\alpha_{2_i}}{\psi_{2_i}}V_{L_i} + \frac{\alpha_{2_i}}{\psi_{2_i}}H_{i,max} + \varphi_i,
\end{equation}
which ensures semi-globally uniformly bounded tracking for the trajectory of \eqref{system_dynamics_1}, i.e.,
\begin{equation*}
\limsup_{t\to\infty}\norm{p_i(t)}  \leq \sqrt{\frac{1}{\psi_{1_i}}(H_{i,max}+\frac{\psi_{2_i}\varphi_i}{\alpha_{2_i}})}.
\end{equation*} 
\end{proof}
\section{Simulation Results}\label{results}
This section presents simulations conducted in MATLAB/Simulink to evaluate the performance of the proposed observer-based control framework for CACC under FDI attacks, disturbance, and bounded measurement noise. Two scenarios are investigated: (i) proposed method without measurement noise, and (ii) proposed method with measurement noise. In all cases, the control objective is to maintain a safe and optimal inter-vehicle distance while mitigating the effects of FDI attack and disturbance. In the simulations, the FDI attack is modeled as a step signal with an initial value of zero, a step time of $t = 30\,\text{s}$, and a magnitude of $0.5$. The desired inter-vehicle distance is set to $5\,\text{m}$. The results include the key performance parameters, the inter-vehicle distance, the estimated position of the leader, and the estimated FDI attack.

All the corresponding system parameters and computed observer gains are provided in Appendix (B).

\subsection{Proposed Method Without Measurement Noise}
In this scenario, measurement noise is neglected to enable a clear evaluation of the proposed interval observer against the nonlinear observer presented in \cite{ansari2024secure}. In the following figures, the original method refers to the approach introduced in \cite{ansari2024secure}, while the baseline method represents the case without attack estimation or mitigation mechanisms.

\begin{figure}[h]
    \centering
    \includegraphics[width=.45\textwidth]{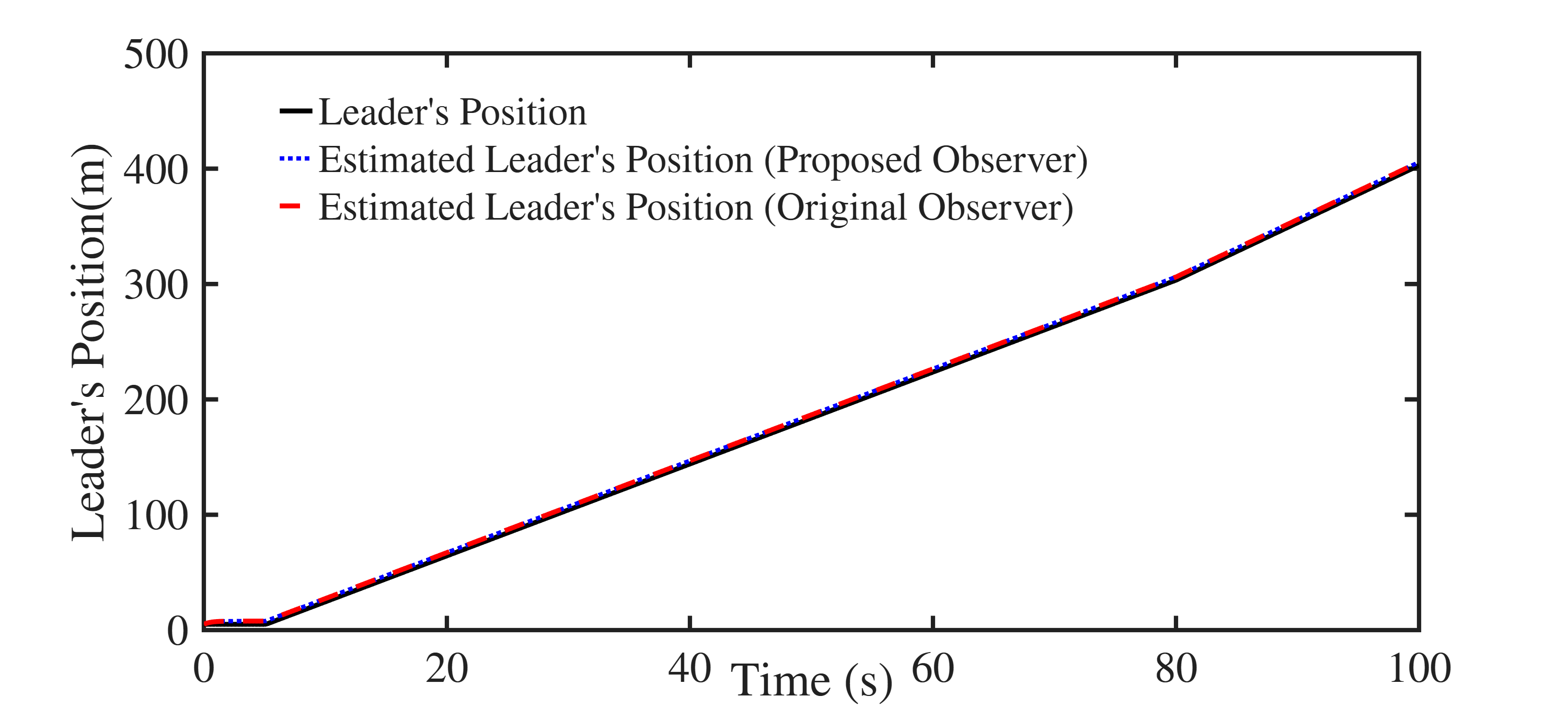}
    \caption{\small{Leader's position estimation.}}
    \label{Leader_Position_without_noise}
\end{figure}

\begin{figure}[h]
    \centering
    \includegraphics[width=.45
    \textwidth]{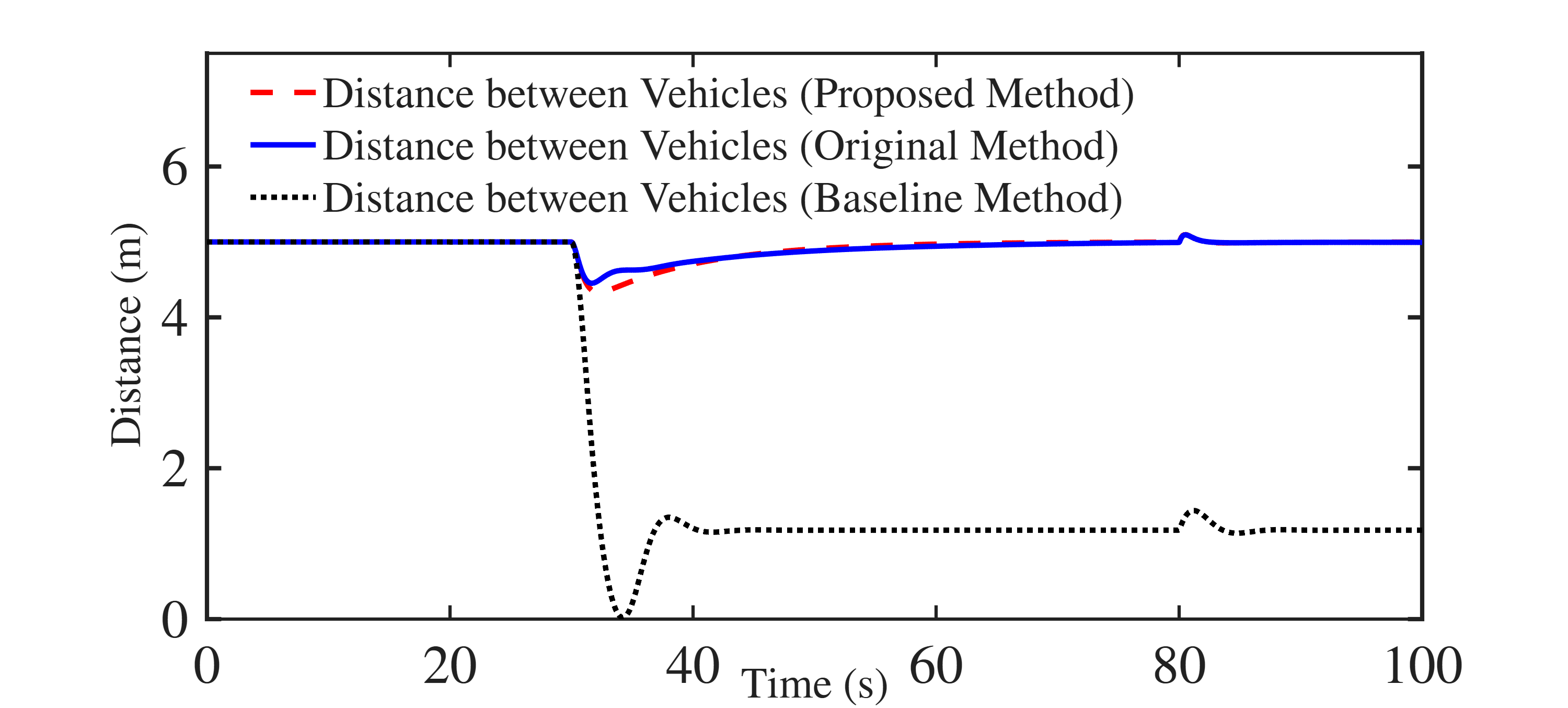}
    \caption{\small{Distance between lead and following vehicles.}}
    \label{Distance_without_noise}
\end{figure}

\begin{figure}[h]
    \centering
    \includegraphics[width=.45\textwidth]{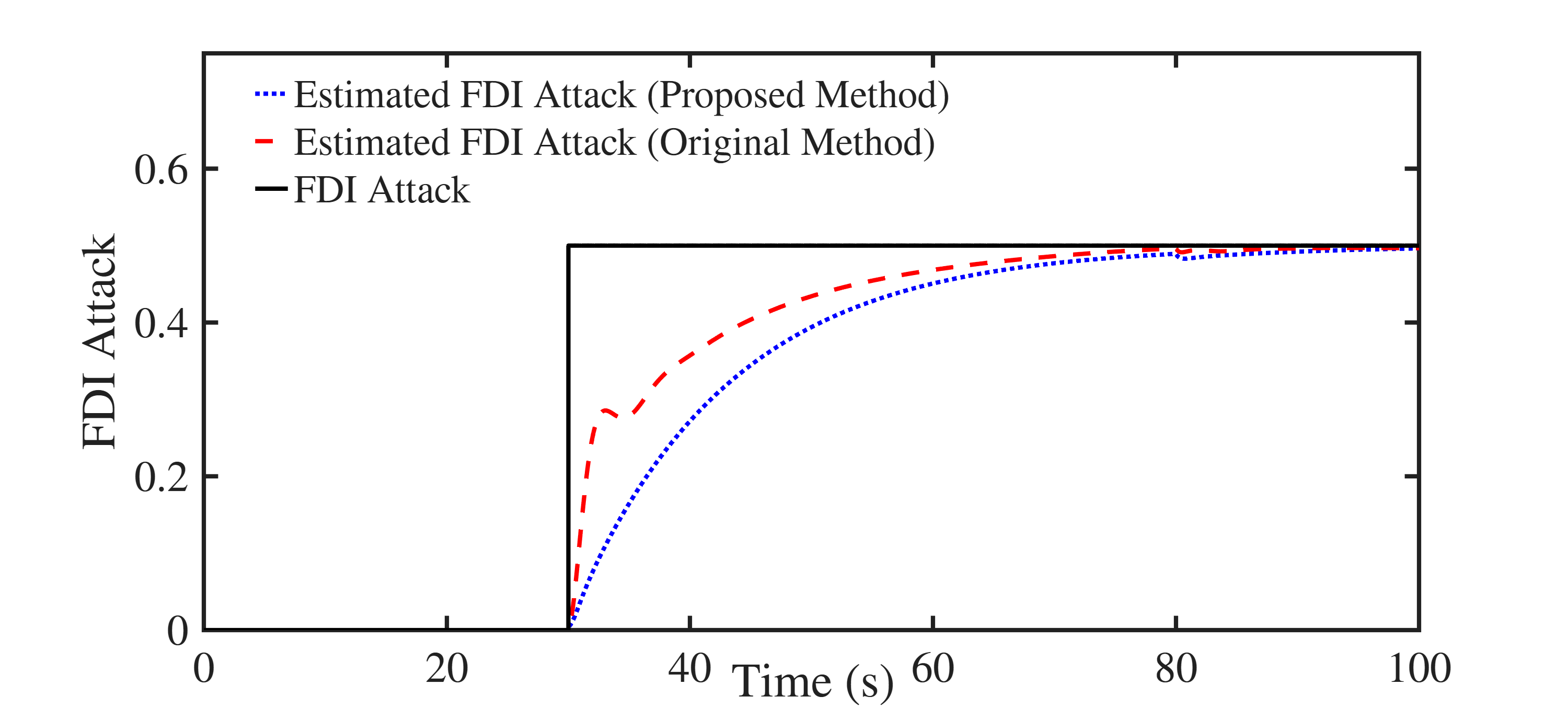}
    \caption{\small{FDI attack estimation}}
    \label{FDI_Estimation_without_noise}
\end{figure}

As shown in Figure~\ref{Leader_Position_without_noise}, both the original observer and the proposed interval observer accurately estimate the leader's position in the absence of measurement noise. However, the original observer achieves slightly better estimation performance under these noise-free conditions. In addition, Figure~\ref{Distance_without_noise} illustrates the inter-vehicle distance for  original, proposed, and baseline methods. Although the original method achieves slightly better tracking performance, both approaches successfully mitigate the effects of the FDI attack occurring at $t = 30\,\text{s}$. Despite a temporary reduction in inter-vehicle distance following the attack, the spacing remains within a safe range. In contrast, the baseline method, which lacks FDI attack estimation and mitigation mechanisms, fails to maintain a safe distance after the attack injection, leading to a potential collision scenario.

The FDI attack estimations obtained from both the original and proposed methods are presented in Figure~\ref{FDI_Estimation_without_noise}. As observed, both methods successfully estimate the injected attack; however, the original method demonstrates slightly better estimation accuracy under noise-free conditions.

\subsection{Proposed Method With Measurement Noise}
To evaluate robustness under more realistic conditions, the same simulation setup is repeated with bounded measurement noise introduced in both position and velocity signals. In practical implementations, sensor readings are inevitably affected by noise; therefore, ensuring observer robustness under such conditions is crucial. In this scenario, the baseline method refers to the case without FDI attack estimation or mitigation mechanisms.
\begin{figure}[h]
    \centering
    \includegraphics[width=.45\textwidth]{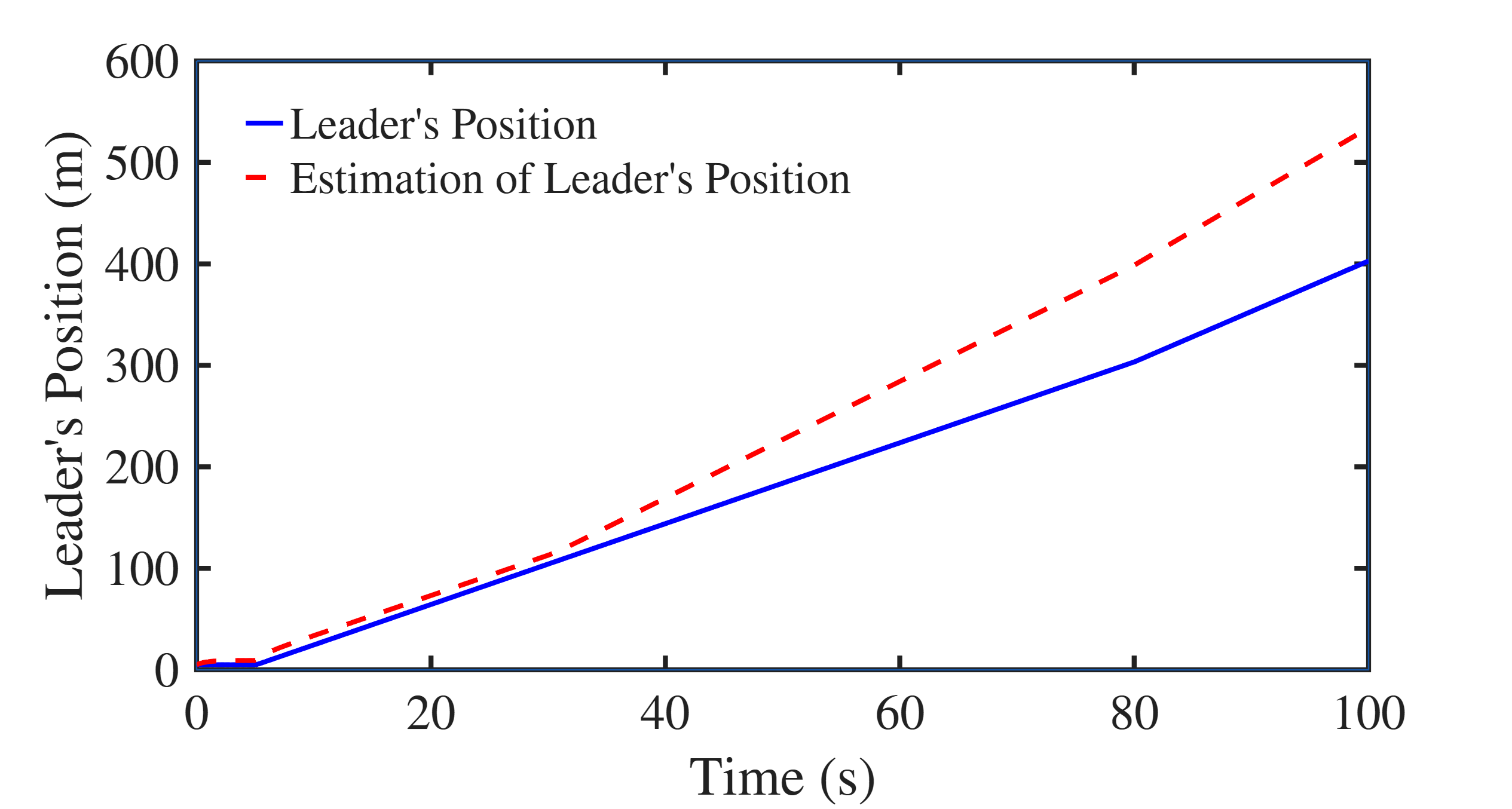}
    \caption{\small{Leader's position estimation}}
    \label{Leader_Position_with_noise}
\end{figure}

As shown in Figure~\ref{Leader_Position_with_noise}, the proposed interval observer explicitly accounts for measurement noise while estimating the leader's position. As expected, the presence of noise introduces slight variations in the estimated trajectory compared to the noise-free case. This scenario is included to emulate real-world conditions, where sensor measurements are inevitably affected by noise. By incorporating noise into the observer design, the proposed method demonstrates its robustness, ensuring reliable position estimation even under noisy measurement environments.

\begin{figure}[h]
    \centering
    \includegraphics[width=.45\textwidth]{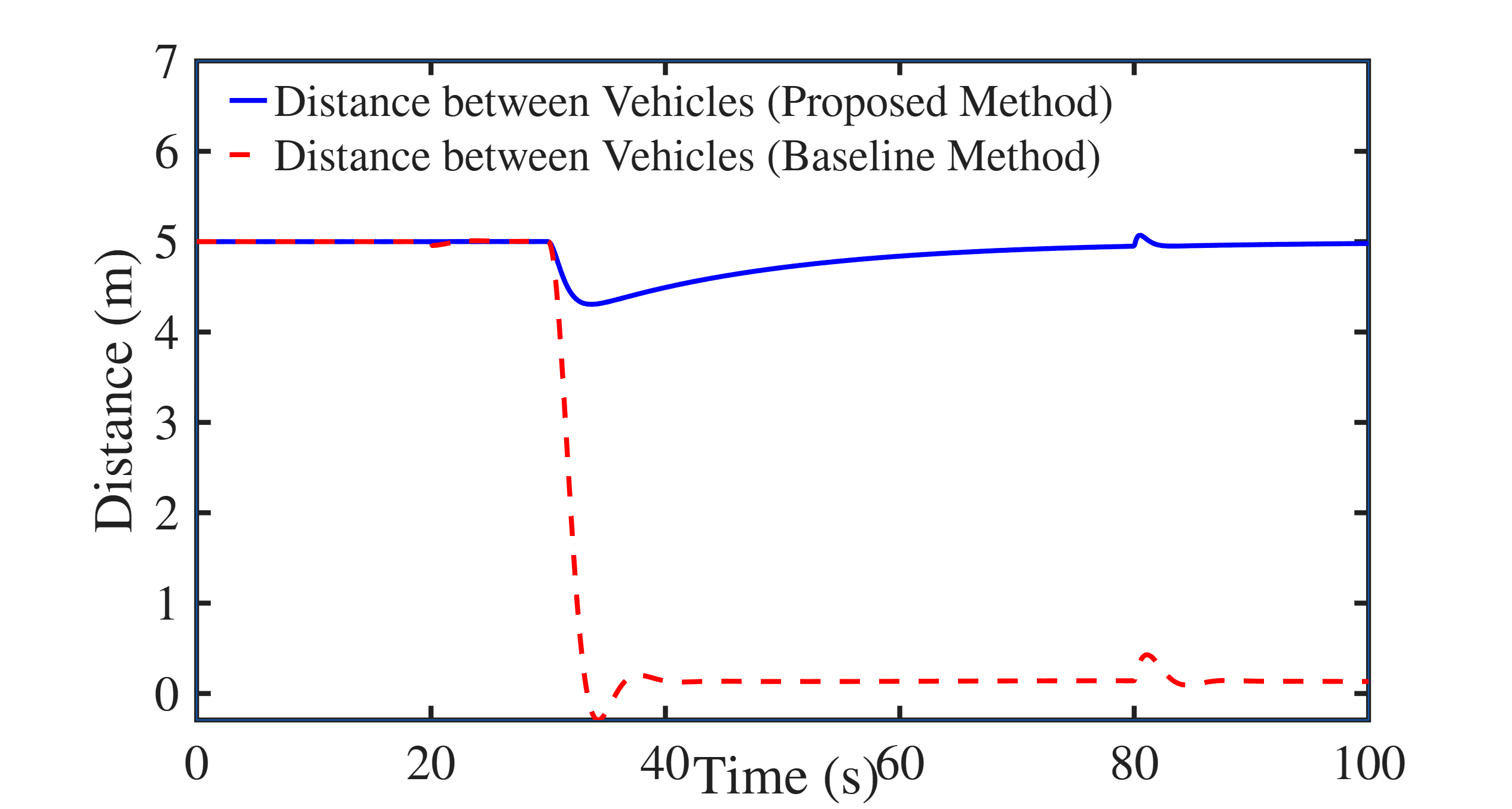}
    \caption{\small{Distance between lead and following vehicles}}
    \label{Distance_with_noise}
\end{figure}

The inter-vehicle distance in the presence of measurement noise, FDI attack, and disturbance is illustrated in Figure~\ref{Distance_with_noise}. Despite slight variations in the estimation of the leader's position, the proposed method maintains acceptable performance in preserving a safe and optimal distance between vehicles under the combined effects of disturbances, FDI attacks, and measurement noise. In contrast, the baseline method fails to sustain the required spacing after the attack injection, resulting in a collision scenario.

\begin{figure}[h]
    \centering
    \includegraphics[width=.45\textwidth]{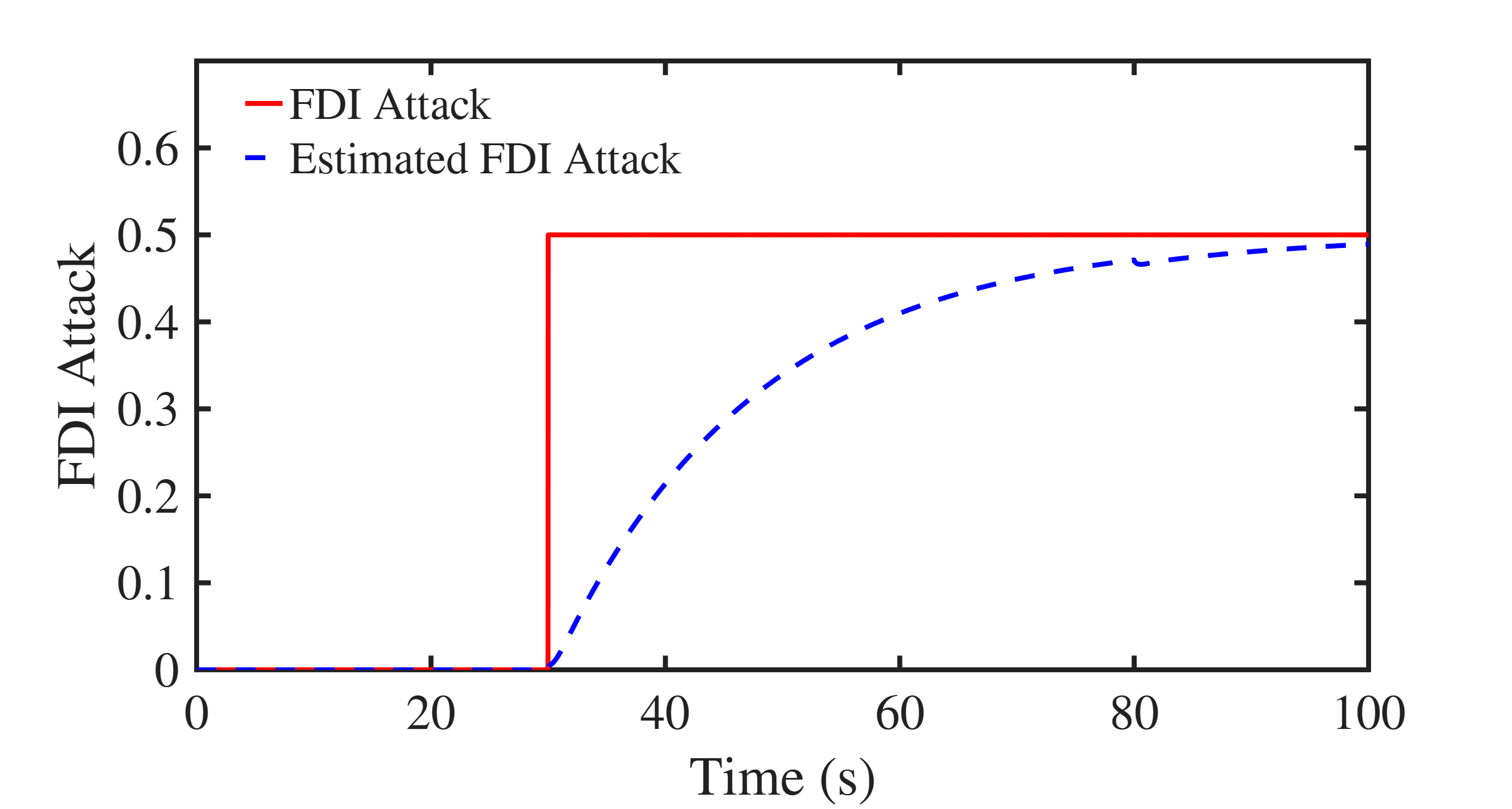}
    \caption{\small{FDI attack estimation}.}
    \label{FDI_Estimation_with_noise}
\end{figure}

Figure~\ref{FDI_Estimation_with_noise} presents the FDI attack estimation obtained using the proposed interval observer in the presence of measurement noise. As observed, despite the influence of noise, the proposed method successfully estimates the injected attack with satisfactory accuracy, demonstrating its robustness under noisy conditions.

To evaluate performance quantitatively, Table \ref{position_estimation} reports the root mean square error (RMSE) between the actual position of the lead vehicle and estimation of it under original observer, proposed interval observer without noise, and proposed interval observer with bounded measurement noise. In addition, RMSE between the actual inter-vehicle distance and the desired distance for all three methods are also presented. As shown in Table \ref{position_estimation}, the RMSE values for both distance and leader’s position are slightly lower in the original method compared to the proposed one, both with and without noise. Nevertheless, the proposed method demonstrates comparable performance and maintains robustness under bounded measurement noise, which is an inevitable factor in real-world scenarios. Therefore, it can be concluded that the proposed interval observer–based framework provides reliable and effective performance under realistic conditions involving FDI attacks, external disturbances, and measurement noise.

\begin{table}[h]
    \centering
    \caption{\small{Root mean square error of estimation of leader's position and actual distance and desired distance}}
    \renewcommand{\arraystretch}{1.4}
    \begin{adjustbox}{max width=\columnwidth,center}
    \begin{tabular}{lccc}
    \hline
     & Original Observer & Interval Observer without Noise & Interval Observer with Noise\\
    \hline
    Position Estimation RMSE & 0.9689 & 2.7873 & 65.2634\\
    Distance RMSE & 0.1356 & 0.2025 & 0.2357\\
    \hline
    \end{tabular}
    \end{adjustbox}
    \label{position_estimation}
\end{table}

\section{Conclusion} \label{s:conclusion}
This paper presented a resilient interval observer-based control framework for CACC under FDI attacks, disturbance, and bounded measurement noise. The proposed approach integrates a nonlinear Lyapunov-based controller with an interval observer that provides guaranteed state bounds for the leade's position from the prespective of follower in the presence of disturbance and bounded measurement noise. A neural network-based estimator was employed to estimate the unknown FDI attack in real time, allowing the controller to effectively mitigate its impact and maintain safe and optimal inter-vehicle spacing. Simulation results demonstrated that the proposed method achieves robust tracking, accurate attack estimation, and resilience against both FDI attack and measurement noise. In noise-free conditions, the interval observer achieves comparable performance to the original nonlinear observer. However, in the bounded measurement noise scenario, the proposed interval–based method naturally exhibits reduced estimation accuracy. This behavior reflects the fact that the method is designed to remain valid under realistic noise bounds, whereas the original point–estimation approach does not accommodate such uncertainties. Even so, the interval observer provides reliable and consistent performance under practical sensing conditions.

\balance
\bibliography{Ref.bib}
\bibliographystyle{ieeetr}
\appendix
\subsection{Neural Network Details}
 The FDI attack, $f_{i-1}$, occurs over a non-compact domain, so a nonlinear mapping, $M_{f_{i-1}}:[t_0,\infty)\to[0,1]$ is required to map time to a compact spatial domain given as
\begin{equation*}
M_{f_{i-1}} \overset{\Delta}{=} \frac{c_{f_{i-1}}(t-t_0)}{c_{f_{i-1}}(t-t_0)+1},\zeta\in[0,1],t\in[t_0,\infty),
\end{equation*}
where $c_{f_{i-1}}\in\mathbb{R}_{>0}$ describes a user-specified gain \cite{chakraborty2017control}. Consequently the FDI attack, $f_{i-1}(t)$, is mapped into the compact domain $\zeta$ as
\begin{equation*}
f_{i-1}(t) = f_{i-1}(M_{f_{i-1}}^{-1}(\zeta)) \overset{\Delta}{=} f_{M_{f_{i-1}}}(\zeta),
\end{equation*}
where $f_{M_{f_{i-1}}}:[0,1]\to\mathbb{R}^{n_i}$ is now defined. The FDI attack can now be estimated using a neural network (NN): 
\begin{equation}\label{neural-network}
\beta_{M_{f_{i-1}}}(\zeta) = W_i^T\sigma(V_i^T\delta_i)+\gamma_i,
\end{equation}
where $\delta_i\in\mathbb{R}^{(n_i+1)\times1}$ signifies the inputs, vectors $W_i\in\mathbb{R}^{(n_i+1)\times n_i}$ and $V_i\in\mathbb{R}^{(n_i+1)\times n_n}$ indicate the unknown ideal weights, and $n_n$ represents the number of neurons in the hidden layer. Additionally, $\sigma$ denotes an activation function vector and $\gamma_i\in\mathbb{R}^{n_i}$ signifies a bounded signal.
Furthermore,
substituting \eqref{neural-network} and \eqref{estimate-NN} into \eqref{estimate-error} yields
\begin{equation*}
\tilde{f}_{i-1} = W_i^T\sigma(V_i^T\delta_i)-\hat{W}_i^T\sigma(\hat{V}_i^T\delta_i)+\gamma_i.
\end{equation*}
Applying Taylor's series approximation: 
\begin{equation}\label{taylor-series}
\tilde{f}_{i-1} = \tilde{W}_i^T\sigma(\hat{V}_i^T\delta_i)+\hat{W}_i^T\sigma'(\hat{V}_i^T\delta_i)\tilde{V}_i^T\delta_i+N_{n_i},
\end{equation}
where
\begin{equation}\label{eq:Nni}
N_{n_i} \overset{\Delta}{=} \tilde{W}_i^T\sigma'(\hat{V}_i^T\delta_i)\tilde{V}_i^T\delta_i+W_i^T\vartheta(\tilde{V}_i^T\delta_i)+\gamma_i,
\end{equation}
and $\tilde{V}_i=V_i-\hat{V}_i$ is the inner NN weight error, $\tilde{W}_i=W_i-\hat{W}_i$ is the outer NN weight error, $\vartheta$ denotes higher order terms, and $N_{n_i}$ is bounded such that $\norm{N_{n_i}}\leq \overline{N}_{n_i}$, with some $\overline{N}_{n_i}>0$.
\subsection{System Parameters \& Computed Gains for the Simulation Example}
\vspace{-.5cm}
\begin{align*}
a_i=0.1413, b_i=6.6870, \alpha_i=1, K_{1_i}=2
\end{align*}
\begin{align*}
\overline{d}=0.01, \underline{d}=-0.01, \overline{\theta}=0.025, \underline{\theta}=-0.025
\end{align*}
For the first scenario: Without Measurement Noise
\begin{align*}
L = 
\begin{bmatrix}
0 \\[2mm]
1.7799
\end{bmatrix}, T =
\begin{bmatrix}
1.0000 & 0 \\[2mm]
0 & -0.0002,
\end{bmatrix}, N =
\begin{bmatrix}
0 \\[2mm]
1.0002
\end{bmatrix}
\end{align*}

For the second scenario: With Measurement Noise
\begin{align*}
L = 
\begin{bmatrix}
0 \\[2mm]
1.0933
\end{bmatrix}, T =
\begin{bmatrix}
1.0000 & 0 \\[2mm]
0 & 0.6244
\end{bmatrix}, N =
\begin{bmatrix}
0 \\[2mm]
0.3756
\end{bmatrix}
\end{align*}
\end{document}